\documentclass[11pt,dvips]{article}
\usepackage{amscd,verbatim}
\usepackage[all]{xy}
\usepackage{graphicx}

\usepackage{amsmath}
\usepackage[psamsfonts]{amssymb}
\usepackage{amsthm}
\usepackage{euscript}

\usepackage{latexsym}

\renewcommand{\(}{\begin{equation}}
\renewcommand{\)}{end{equation} \vspace{-.05in}\linebreak}

\newcounter{saveeqn}
\newcounter{savealpheqn}

\newcommand{\alpheqn}{\setcounter{saveeqn}{\value{equation}}%
 \stepcounter{saveeqn}\setcounter{equation}{0}%
 \renewcommand{\theequation}{\mbox{\arabic{section}.\arabic{saveeqn}
\alph{equation}}}
 \renewcommand{\)}{\end{equation}}}
\def\part#1{\frac{\partial}{\partial{#1}}}%
\def\group#1{\refstepcounter{equation}\setcounter{saveeqn}{\value{equation}}%
 \label{#1}\setcounter{equation}{0}%
\renewcommand{\theequation}{\mbox{\arabic{section}.\arabic{saveeqn}
\alph{equation}}}
 \renewcommand{\)}{\end{equation}}}
\newcommand{\reseteqn}{\setcounter{equation}{\value{saveeqn}}%
 \renewcommand{\theequation}{\arabic{section}.\arabic{equation}}%
 \renewcommand{\)}{\end{equation}}}

\newcommand{\aalpheqn}{\setcounter{saveeqn}{\value{equation}}%
 \stepcounter{saveeqn}\setcounter{equation}{0}%
 \renewcommand{\theequation}{\mbox{
       \Alph{subsection}.\arabic{saveeqn}\alph{equation}}}
  \renewcommand{\)}{\end{equation}}}
\newcommand{\areseteqn}{\setcounter{equation}{\value{saveeqn}}%
 \renewcommand{\theequation}{\Alph{subsection}.\arabic{equation}}%
 \renewcommand{\)}{\end{equation}}}

\renewcommand{\thefootnote}{\alph{footnote}}
\renewcommand{\(}{\begin{equation}}
\renewcommand{\)}{\end{equation}}
\newcommand{\ba}{\begin{eqnarray}}
\newcommand{\ea}{\end{eqnarray}}

\newcommand{\bp}{\mathop{\vtop{\ialign{##\crcr
  $\hfil\displaystyle{}\hfil$\crcr\noalign{\kern-13pt\nointerlineskip}
  \BIG{(}\hskip0pt\crcr\noalign{\kern3pt}}}}}
\newcommand{\cbp}{\mathop{\vtop{\ialign{##\crcr
  $\hfil\displaystyle{}\hfil$\crcr\noalign{\kern-13pt\nointerlineskip}
  \BIG{)}\hskip0pt\crcr\noalign{\kern3pt}}}}}
\newcommand{\pa}{\mathop{\vtop{\ialign{##\crcr
  $\hfil\displaystyle{\oplus}\hfil$\crcr\noalign{\kern+1pt\nointerlineskip}
  \hspace{.08in}$^{\alpha=0}$\hskip6pt\crcr\noalign{\kern3pt}}}}}
\renewcommand{\sp}{,\hspace{.3in}}
\newcommand{\p}{^\prime}

\newcommand{\R}{\ensuremath{\mathbb R}}

\newcommand{\Z}{\ensuremath{\mathbb Z}}

\def\dwn{\downarrow}

\newcommand{\beq}{\begin{equation}}
\newcommand{\eeq}{\end{equation}}

\numberwithin{equation}{section}

\def\hsp#1{\hspace{#1in}}

\catcode`\@=11
\def\vereq#1#2{\lower3pt\vbox{\baselineskip1.5pt \lineskip1.5pt
\ialign{$\m@th#1\hfill##\hfil$\crcr#2\crcr\sim\crcr}}}
\catcode`\@=12

\makeatletter
\newcommand\figcaption{\def\@captype{figure}\caption}
\newcommand\tabcaption{\def\@captype{table}\caption}
\makeatother
\renewcommand{\(}{\begin{equation}}
\renewcommand{\)}{\end{equation}}

\newcommand{\RR}{{\mathbb R}}

\theoremstyle{plain}

\theoremstyle{definition}

\newcommand{\twoa}{\text{I}\!\text{IA}}
\newcommand{\twob}{\text{I}\!\text{IB}}

\begin{document}

\begin{titlepage}
\begin{flushright}
IFUP-TH/2003/47

hep-th/0311235
\end{flushright}

\vspace{2em}
\def\thefootnote{\fnsymbol{footnote}}

\begin{center}
{\Large\bf From $E_8$ to $F$ via $T$}
\end{center}
\vspace{1em}

\begin{center}
Jarah Evslin\footnote{E-Mail: jarah@df.unipi.it}\ 
\end{center}

\begin{center}
\vspace{1em}
\hsp{.3}\\
{\em INFN Sezione di Pisa\\
     Via Buonarroti, 2, Ed.~C,\\
     56127 Pisa, Italy}\\

\end{center}

\vspace{0em}
\begin{abstract}
\noindent
We argue that T-duality and F-theory appear automatically in the $E_8$ ``gauge'' bundle perspective of M-theory.  The 11-dimensional supergravity four-form determines an $E_8$ bundle.  If we compactify on a two-torus, this data specifies an $LLE_8$ bundle where $LG$ is a centrally-extended loopgroup of $G$.  If one of the circles of the torus is smaller than $\sqrt{\alpha\p}$ then it is also smaller than a nontrivial circle $S$ in the $LLE_8$ fiber and so a dimensional reduction on the total space of the bundle is not valid.  We conjecture that $S$ is the circle on which the T-dual type \twob\ supergravity is compactified, with the aforementioned torus playing the role of the F-theory torus.  As tests we reproduce the known T-dualities between NS5-branes and KK-monopoles, as well as D6 and D7-branes where we find the desired F-theory monodromy.  Using Hull's proposal for massive \twoa, this realization of T-duality allows us to confirm that the Romans mass is the central extension of our $LE_8$.  In addition this construction immediately reproduces the conjectured formula for global topology change from T-duality with $H$-flux.
\end{abstract}

\vfill

\today 

\end{titlepage}
\setcounter{footnote}{0} 
\renewcommand{\thefootnote}{\arabic{footnote}}

\pagebreak
\renewcommand{\thepage}{\arabic{page}}

\section{Introduction}
The low energy limit of M-theory is eleven-dimensional supergravity.  Classically eleven-dimensional supergravity is unique.  However off-shell, if we relax manifest Lorentz-invariance, distinct variants exist \cite{NdW,Nic}.  The manifestly Lorentz-invariant variety commonly featured in the literature is responsible for much of what we know of M-theory, from the BPS soliton spectrum to the embeddings of supersymmetric gauge theories.  Yet it has a number of limitations which may lead one to search for an alternate low energy description.  We shall now describe three features of M-theory that this SUGRA formulation does not capture.

Perhaps the most obvious is that it misses the $E_8$ super Yang-Mills which inhabits every boundary component of space-time \cite{HW}, instead these must be added by hand.  In particular if M-theory is compactified on a 4-manifold $M$ crossed with six-dimensional Minkowski space crossed with an interval one finds an $E_8$ gauge bundle on each 10-dimensional end of the interval.  These $E_8$ bundles are topologically specified entirely by their instanton numbers $I_1$ and $I_2$ on the two ends of the world.  The ten-dimensional vector multiplet contains a Weyl spinor and one must choose whether the spinors on the two ends have the same chirality as in Fabinger-Ho$\check{\textup{r}}$ava \cite{FH} or opposite chirality as in Ho$\check{\textup{r}}$ava-Witten \cite{HW}.  In these two cases one finds that the instanton numbers obey (see for example Ref.~\cite{DFM})
\begin{eqnarray}
I_1=I_2 && \textup{in the Fabinger-Ho$\check{\textup{r}}$ava case}\nonumber\\
I_1+I_2=-\int_M\frac{p_1(M)}{2} && \textup{in the Ho$\check{\textup{r}}$ava-Witten case}
\end{eqnarray}
where $p_1(M)$ is the first Pontrjagin class of $M$.

The second limitation arises from the fact that T-duality is obscured in 10$d$ SUGRAs as the stringy modes are not included in the low energy effective theory.  11-dimensional supergravity compactified on a 2-torus yields type \twoa\ $10d$ SUGRA compactified on a circle.  This is the low energy effective theory of \twoa\ string theory, and so when the radius $R$ of this circle is smaller than $\sqrt{\alpha\p}$ the \twoa\ description must break down in favor of an equivalent dual \twob\ description compactified on a circle of radius $\alpha\p/R$.  This transformation takes NS5-branes to KK-monopoles and more generally takes $H$-flux on the \twoa-side to curvature of the circle bundle on the \twob-side as has been quantified in \cite{BEM}.  The \twob\ theory comes with an elliptic fibration, where the fibers transform nontrivially as one encircles a D7-brane.  A further T-duality transverse to this D7-brane yields a D8-brane in Romans' massive \twoa, a deformation of \twoa\ supergravity not evident in the original 11-dimensional description.  

Dirac quantization in abelian gauge theory is a consequence of the existence of the gauge bundle, which must have an integral Chern class.  In 11$d$ SUGRA, without mentioning 2-gerbes, such a topological explanation for the quantization of the four-form field strength is lacking.  In particular the twisted quantization condition \cite{FluxQuant}
\begin{equation}
G_4+\frac{p_1(M)}{4}\in\Z
\end{equation}
which holds for M-theory on any orientable $spin$ manifold \cite{FluxQuant,BEM} is mysterious.

In the future a formulation of 11$d$ SUGRA may be found in which the above M-theory features are manifest.  As a possible preliminary step to such a program, in this note we will follow Refs.~\cite{FluxQuant,DMW,GM,uday,allan,Morrison} and introduce an $E_8$ ``gauge'' bundle over the 11$d$ bulk of 11$d$ supergravity such that
\begin{equation}
G_4=\frac{2Tr F^2+Tr R^2}{16\pi} \label{const}
\end{equation}
where $F$ and $R$ are the curvatures of the gauge bundle and the tangent bundle respectively.  Of course one may always include such an auxiliary bundle as a bookkeeping device, and we will not address the critical question of what kind of dynamics it may have, if it has any dynamics at all.  For example the bundle may be purely topological with propagating degrees of freedom appearing only holographically on the end of the world as in the proposal of Ref.~\cite{DFM}.  Instead we will consider the more modest goal of motivating the introduction of this bundle by using it to reproduce the above M-theory facts in Secs.~\ref{msec} and \ref{twosec}.  Note for example that the final paragraph, on the quantization condition, is a trivial consequence of the construction (\ref{const}), the existence of the $E_8$ bundle, and the fact that characteristic classes are integral. 

T-duality will be seen to be a choice of which circle is considered to be part of space-time and which is considered to be part of a fiber over space-time.  To confirm that this is in fact T-duality we will work through several examples in Sec.~\ref{exsec}.  In Section~\ref{romsec} we will use Hull's proposal, which equates massive \twoa\ string theory with M-theory compactified on a nilmanifold, to show that our claim implies an older conjecture \cite{allan} that massive \twoa\ naturally comes with a loop group of $E_8$ bundle centrally-extended by the Romans mass. 

Several other applications of such $E_8$'s have been known for quite some time.  For example Diaconescu, Moore and Witten have used this $E_8$ bundle to calculate a topological phase in the M-theory partition function \cite{DMW,MS}.  In fact it is only with this bundle's aid that the phase has been shown to be well-defined \cite{FluxQuant}.  It was also hinted that such a construction may be useful for the classification of orientifold planes in Ref.~\cite{Kentaro}.

A much more mysterious application predates M-theory.  The dimensional reduction of 11$d$ SUGRA on $AdS^4\times S^7$ is only equivalent to four-dimensional gauged supergravity after using the classical equations of motion.  This means that the quantum version, whatever it may be, of the Lorentz-invariant 11$d$ SUGRA does not reduce to the quantum 4$d$ SUGRA.  Clearly we would like M-theory to reduce successfully to the quantum gauged 4$d$ SUGRA.  There is an 11d SUGRA in which this dimensional reduction succeeds even off-shell, this is the variant of De Wit and Nicolai \cite{NdW,Nic} in which the structure group is enhanced to $E_{8,8}$ via the inclusion of extra degrees of freedom which are pure gauge.  The relation of this $E_{8,8}$ to our compact $E_8$ is quite mysterious, although one might guess that at least the maximally compact $SO(16)$'s are the same.

\section{M-Theory and Heterotic Solitons from $E_8$} \label{msec}
\subsection{Soliton Spectra from Homotopy Groups} \label{hom}
Many times throughout this paper we will need to compute the soliton spectrum\footnote{Here the term ``soliton'' is thoroughly abused to include anything which may be electrically or magnetically charged under $G$.} of a theory with gauge group $G$.  This subsection contains a brief summary of the usual construction of $G$-solitons from defects in $G$ bundles.  The reader familiar with this construction may wish to skip to Subsection~\ref{msub}.

We may construct any $G$ bundle over $S^{k+1}$ as follows.  The $G$ bundle may be trivialized over the northern and southern hemispheres $S^{k+1}_N$ and $S^{k+1}_S$.  It is then determined entirely by the transition function, which is a map $f$ from the equatorial $S^k$ to $G$ and so is classified up to homotopy by $\pi_k(G)$.  Here we recall that $\pi_k(G)$, the $k$th homotopy group of $G$, is a group of maps from the $k$-sphere to $G$, that is $f:S^k\longrightarrow G$.  Homotopic maps are identified to the same group element.  In this note, except for the case of massive \twoa, the relevant homotopy groups will be $\pi_k(G)=\Z$.  In these cases our bundles are also characterized by a $k+1$-form characteristic class $\chi_{k+1}$ such that
\begin{equation}
\int_{S^{k+1}}\chi_{k+1}=[f]\in\pi_k(G)=\Z. \label{gauss}
\end{equation}

Consider a space-time which contains a contractible $(k+1)$-sphere.  $G$ bundles over this sphere are parametrized by integers, as in Eq.~(\ref{gauss}).  However the sphere may be contracted away, and so by Stoke's theorem the integral of $d\chi_{k+1}$ over the interior of the sphere must be equal to $[f]\in\Z$.  The fact that $[f]$ is integral for any sphere chosen means that integrals of $d\chi_{k+1}$ must also be integral, and so $d\chi_{k+1}$ is the sum of the Dirac delta functions supported on codimension $k+1$ submanifolds.  At each of these delta functions $G$ is not defined, and so these submanifolds are defects in the bundle.  We say that defects constructed in this way are magnetically charged under $\chi_{k+1}$ and we interpret Eq.~(\ref{gauss}) as Gauss' Law for measuring the magnetic charge linked by a $(k+1)$-sphere.  These defects intersect the $(k+2)$-dimensional interior of the $(k+1)$-sphere at a discrete set of points and so they must be codimension $(k+2)$, therefore the linking number with $S^{k+1}$ may be defined.

In addition to magnetic solitons there are electric solitons whose charge we define with the Hodge-dual Gauss' Law
\begin{equation}
Q=\int_{S^{d-k-1}}*\chi_{k+1}\in\Z
\end{equation}
where $d$ is the dimension of space-time and $*$ is contraction with the epsilon tensor.  $S^{d-k-1}$ is a sphere that links an electric soliton, and so these solitons must be $k$-dimensional.  In ordinary gauge theories purely electrically-charged objects do not correspond to nontrivial bundle configurations and must be added to the theory by hand.  However in M-theory and string theories, due to Chern-Simons terms present in the low energy effective actions, it will be seen that magnetic solitons may carry electric charge and so electric solitons are automatically included in the classification of $G$ bundles.

\subsection{M-Theory Solitons} \label{msub}
The above construction may be straightforwardly applied to classify the solitons of an $E_8$ bundle over an 11-manifold $M^{11}$.  $E_8$ is a simple Lie group and so $\pi_3(E_8)=\Z$, however it is unique among semi-simple Lie groups in that every other homotopy group up through $\pi_{14}$ is trivial.  This means that the only magnetic solitons on an 11-manifold will be those arising from $\pi_3$.  These are the codimension 5 solitons whose charge is defined, via Gauss' Law, by integrating the characteristic class
\begin{equation}
\chi_4=G_4-\frac{p_1(M)}{4}
\end{equation}
over a linking 4-sphere.  The division of $p_1(M)$ by $4$ is canonically defined if M-theory is compactified on a $spin$ manifold, as we will assume throughout this note.  The image of the nontrivial map from $S^3$ to $E_8$ is homotopic to an $SU(2)$ in the $E_8$, and so this construction remains valid if the $E_8$ gauge symmetry is broken so long as a nonabelian factor remains in which to embed this $SU(2)$.  If it is broken still further, a nonabelian factor must be restored in the region of this defect, that is at least one Higgs field that breaks an $SU(2)$ must vanish in the defect's core. 

If we interpret $G_4$ as the 4-form field strength of 11$d$ SUGRA, then these magnetic solitons are M5-branes \cite{GM}.  The shift by the first Pontrjagin class of the tangent bundle, $p_1(M)$, is responsible for the shifted quantization condition which leads, for example, to the half-integral $G_4$ charges of the OM5-plane.  Notice that any other semi-simple group would have additional nontrivial low-dimensional homotopy classes and thus additional magnetic solitons beyond the known soliton spectrum of 11$d$ SUGRA.

One may imagine that 11$d$ SUGRA is somehow the dimensional reduction of the total space of this $E_8$ bundle, with many degrees of freedom missing or gauged away.  Motivated by this perspective one may formally define the ``size'' of the $E_8$ fiber over each point in space-time.  If we wish to make our M5-branes nonsingular, the $E_8$ fiber, or more specifically the $SU(2)$ subgroup which is the image of the transition function $f$, should degenerate over the M5 just as the circle fiber degenerates at a Kaluza-Klein monopole.  Of course in the classical supergravity solution M5-branes live at the end of an infinite throat and so the fiber never vanishes altogether.  

There is only one length scale in the theory, the Planck length $l_p$, and we define the size of the $SU(2)$ to be $l_p^3$.  The fact that the $SU(2)$ degenerates around the M5-brane suggests that the tension of the M5-brane depends entirely on $l_p$, and so by dimensional analysis is proportional to $l_p^{-6}$, by analogy with the calculation of the tension of the KK-monopole.  This interpretation has two attractive features.  First, by having the effective Plank scale shrink to zero around the M5-brane, so long as it shrinks quickly enough, the M5-brane is naturally at the end of a throat which is infinite in Planck units, in accordance with the classical solution.  Secondly, at scales of order the Plank scale the 11-dimensional picture breaks down as expected, this breakdown is a result of the KK-modes of the $E_8$ bundle which, as we will see when we consider a dimensional reduction, are M2-branes.

We have found the objects magnetically charged under $G_4$, the M5-branes.  At this point we may, following the usual logic in gauge theories, introduce M2-branes by hand as electrically-charged matter.  However M2-branes are already included due to a miracle of the Chern-Simons term $C_3\wedge G_4\wedge G_4$ in the 11-dimensional SUGRA action.  This term ensures that an M5-brane wrapping the submanifold $N^6\subset M^{11}$, being Poincare dual to $dG_4$, has a worldvolume action containing the terms
\begin{equation}
S_{M5}\supset\int_{M^{11}}G_4\wedge *G_4 + C_3\wedge G_4\wedge G_4 = \int_{M^{11}}dG_4\wedge (C_6+C_3\wedge C_3)=\int_{N^6}C_6+C_3\wedge C_3
\end{equation}
where $dC_6=*G_4$.  The first term of the expression on the right indicates that M5-branes are electrically charged under $C_6$.  The second implies that in the presence of $C_3$ flux M5-branes also carry electric charge under $C_3$, which is M2-brane charge.  For example if an M5-brane wraps $N^6=\R^3\times S^3$ where $S^3$ bounds a 4-ball $B^4$ such that $\int_{B^4}G_4=k$ then $\int_{S^3} C_3=k$ and so the M5-brane carries $k$ units of M2-brane charge, and in fact may decay into $k$ M2-branes extended along the $\R^3$.

\begin{figure}[ht]
  \centering \includegraphics[width=3in]{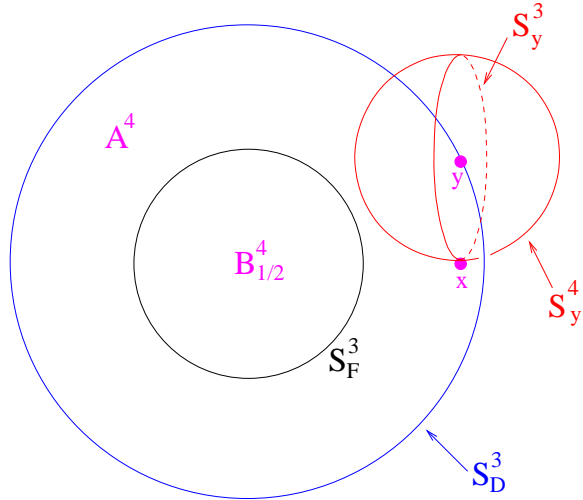}
\caption{An M5-brane wraps a trivial 3-cycle $S^3_D$.  This leads to a nontrivial $E_8$ bundle over every linking $S^4_y$, characterized by a transition function $f_y:S^3_y\longrightarrow E_8$.  This transition function represents a nontrivial class in $\pi_3(E_8)$.  A choice of $x\in S^3_y$ in each $S^4_y$ is mapped to some basepoint in $E_8$.  $S^4_y$'s are labeled by their centers $y\in S^3_D$, and thus for each $x$ there is a map $S^3_D\longrightarrow E_8:y\mapsto f_y(x)$.  $E_8$ approximates the classifying space $K(\Z,3)$ and therefore the homotopy class of the map determines an element of the 3rd cohomology group of the M5-brane's worldvolume.  This element is the worldvolume 3-form field strength $T_3$, which is related to $C_3$ by a gauge transformation, and so the homotopy class of this map equals the M2-brane charge $\int_{S^3_D} T_3$ of the dielectric M5-brane.} \label{m2}
\end{figure}

In SUGRA M5-branes may carry M2-brane charge as a result of the Chern-Simons term.  We will now see that this effect of the Chern-Simons term is automatically built into the topology of the $E_8$ bundle.  The above SUGRA construction of an M2-brane may be translated to a topological construction of the M2-brane as a defect in the $E_8$ bundle \cite{Feb}.  Following the previous paragraph, we begin with an M5-brane which wraps $\R^3$ crossed with a contractible $S^3_D$, as depicted in Fig.~\ref{m2}.  This M5-brane is the codimension 5 magnetic defect of the $E_8$ bundle, and so any 4-sphere $S^4_y$ that links a point $y\in S^3_D$ supports a nontrivial $E_8$ bundle with characteristic four-class equal to one.  This non-trivial bundle may be trivialized over the northern and southern hemispheres of the 4-sphere, and so is characterized entirely by the transition function on the equatorial three-sphere $S^3_y$.   The transition function is a map $f_y:S^3_y\longrightarrow E_8$ and so is classified by $\pi_3(E_8)=\Z$.  The M5-brane charge at the point $y$ is entirely specified by the homotopy class of this map because any map in its homotopy class may, by definition, be continuously deformed to any other such map. 

While this deformation is possible at any single value of $y$, there is a global obstruction to performing such a deformation at every $y$ simultaneously.  This is because globally the bundle carries yet another topological invariant besides the homotopy class of $f_y$.  To see this other invariant we continuously choose\footnote{Different choices correspond to gauge transformations of $C_3$.  As we will see, large gauge transformations \textit{can} change the M2-brane change.} a point $x$ on each 3-sphere $S^3_y$.  Any such gauge choice yields a continuous map
\begin{equation}
g_x:S^3\longrightarrow E_8:y\mapsto f_y(x)
\end{equation} 
of basepoints of our transition function.  This map, like $f_y$, is topologically classified by $\pi_3(E_8)=\Z$ and so provides the second invariant.  

Using the above SUGRA construction of dielectric M2-branes, we may see that this new integral invariant is M2-brane charge by showing that it is equal to the amount of $G_4$ flux on a 4-disk bounded by $S^3_D$.  This corresponds to a nontrivial bundle on the 4-disk, which may be constructed by partitioning the 4-disk into a core 4-disk which is surrounded by a 4-annulus.  We will name the 3-sphere which separates the core from the annulus $S^3_F$.  The bundle over the 4-disk is then characterized by the transition function $h:S^3_F\longrightarrow E_8$ which, by our interpretation of $G_4$-flux,\footnote{Here we are assuming that the $p_1(M)$ contribution may be ignored.  This would not be possible, for example, if $S^3_D$ linked an OM5-plane. This transition function is an example of the large gauge transformations mentioned in the previous footnote.} corresponds to the element $[h]=k\in\pi_3(E_8)$.  If we shrink $S^3_D$ so that it lies within the core then $g$ is multiplied by $h$ and so the homotopy class of $g$ is augmented by the homotopy class of $h$.  Thus the homotopy class of $g$ shifts by $k$ as the M5 sweeps out $k$ units of $G_4$ flux, which means that $C_3$ changes by $k$ units.  As described above, $C_3$ measures the M2-charge and so the homotopy class of $g$ is equal to the M2 charge plus a constant.  

This seems to be the result that we sought, except that the integral of $C_3$ need not be integral and there is an undetermined constant.  Of course the homotopy class of $g_x$ is automatically integral and so the constant shift must save the integrality.  This constant arises from M2-brane charge contributions from the bulk \cite{Wati}.  When these are included one finds that the charge is always integral and in fact is equal to the integral of $T_3$, the self-dual 3-form field strength which inhabits the worldvolume of the M5-brane.  $C_3+T_3$ is the only gauge invariant quantity and when the integral of $C_3$ is an integer we may gauge transform it entirely into $T_3$.  However there are other gauges in which $T_3$ integrates to a different value, and so the M2-brane charge is not gauge-invariant.  

For example, imagine that space-time contains a non-contractible 4-sphere with $k$ units of $G_4$-flux.  Then $C_3$ cannot be globally defined.  However we may define it patchwise on the northern and southern hemispheres.  We may then wrap an M5-brane around the equator $S^3_D$.  One may think of this M5-brane as an inhabitant of the northern or the southern patch, but these two answers give a different $\int_{S^3_D}C_3$ by $k$ units.  $C_3+T_3$ is gauge invariant and so the integrals of $T_3$ and therefore the M2-charges must also differ by $k$ units.  In fact, the M2-charge in such a background is only defined moduli $k$ as a result of a MMS instanton \cite{MMS} in which $k$ M2-branes grow into an M5-brane which sweeps out the 4-sphere and then annihilates itself. The fact that this M5-brane loses its M2-charge is, if we replace the 4-sphere by $S^3\times S^1$, an M-theory lift of a Freed-Witten anomaly.  Thus we see that this particular Freed-Witten anomaly is built into the $E_8$ bundle formalism.  For a less trivial example of a Freed-Witten anomaly manifested as an obstruction to the existence of the $E_8$ bundle see Ref.~\cite{slow}.

We have now seen that the $E_8$ bundle picture knows that M2-brane charge is not gauge-invariant, but rather may be gauge-transformed into integral $C_3$-flux.  As a result, to discuss the tension of an M2-brane we will need to fix a gauge.  To do this we will consider a dielectric brane which is an M5-brane wrapping a cycle $S^3_D$ which is much smaller than the scales of any fluxes or nontrivial cycles in the theory, and then we will take its size to zero.  A natural gauge is thus one in which $C_3$ is well defined in the interior of $S^3_D$, and so its integral over $S^3_D$ vanishes as the sphere is shrunk to zero.

Naively the tension of the M2-brane is just the tension of the M5-brane times the volume of the cycle that it wraps.  In the spirit of considering a reduction from the total space of the bundle, we will formally define the volume of this cycle in the total space of the $E_8$ bundle so that this intuition holds.  In the total space an M5-brane containing a single unit of M2-brane charge has a transition function $g_x$ which for any choice of basepoint $x$ wraps an $SU(2)$ in the $E_8$ once.  We will interpret this as the M5 ``wrapping'' the $SU(2)$ once, and thus the volume of the M5 along the 3 compact directions is the volume of this $SU(2)$, which we have defined to be $l_p^3$.  This gives an M2 tension that is equal to $l_p^3$ times the M5-brane tension, or $l_p^{-3}$.  While the result is correct, this tension calculation is a very weak test of our viewpoint as $l_p$ is the only dimensionful quantity in our theory and so all tensions are already determined by dimensional analysis.  When we turn our attention to type \twoa\ this will no longer be the case.    

\subsection{Heterotic Solitons}
The ordinary formulation of 11$d$ SUGRA does not provide an origin for the $E_8$ bundles that inhabit each 10-dimensional boundary component of space-time.  Any 11$d$ $E_8$ bundle formulation is faced with the opposite problem, it must explain why the bulk $E_8$ gauge bosons do not appear to couple perturbatively to the familiar subset of the M-theory degrees of freedom.  While this question is quite difficult to answer, a boundary $E_8$ bundle is easily produced in such a formulation.  A 10$d$ $E_8$ gauge bundle on each boundary component may be constructed simply by restricting the 11$d$ $E_8$ gauge bundle to the boundary.  The reduced supersymmetry on the boundary allows the gauge bosons to interact with the boundary degrees of freedom, and indeed each boundary component contains a Weyl fermion charged under this $E_8$.  While this coupling is clearly allowed by 10$d$ $N=1$ SUSY on each boundary, the fact that it is actually manifested must be shown to be a consequence of any proposed dynamics for this $E_8$ bundle, as has been shown in the proposal of Ref.~\cite{DFM} for Fabinger-Ho$\check{\textup{r}}$ava compactifications.

The restriction of the 11$d$ bundle to 10$d$ boundaries clearly gives an $E_8$ bundle, but one may wonder whether it gives the right $E_8$ bundle.  In a sense this is trivial, as the construction (\ref{const}) appears to be the known anomaly cancellation condition for boundary fields from Ref.~\cite{HW}.  Of course this construction extends that relation to the bulk\footnote{This extension to the bulk exists and is unique \cite{Stong}.}, but the fact that it agrees with the known relation on the boundary means that it must give the right boundary bundle.  

This picture is complicated by the choice of chirality of the Weyl spinors on each boundary component.  To see this, we will restrict our attention to M-theory compactifications on $M^4\times \R^6\times I$ where $M^4$ is a compact 4-manifold and $I$ the interval.  This space-time has two boundary components, which are copies of $M^4\times \R^6$.  If the same chirality is chosen on both components then Eq.~(\ref{const}) is indeed the boundary anomaly-cancellation condition.  We may then use Stoke's theorem to count the difference between the instanton numbers $I_1$ and $I_2$ on the two boundaries
\begin{eqnarray}
I_1-I_2&=&\frac{1}{16\pi^2}(\int_{(M^4,0)}Tr(F^2)-  \int_{(M^4,1)}Tr(F^2))\nonumber\\
&=&\frac{1}{32\pi^2}(\int_{(M^4,0)}(2Tr(F^2)+Tr(R^2))-  \int_{(M^4,1)}(2Tr(F^2)+Tr(R^2)))\nonumber\\
&=&\frac{1}{2\pi}(\int_{(M^4,0)}G_4-  \int_{(M^4,1)}G_4)=\frac{1}{2\pi}\int_{M^4\times I} dG_4=Q_{M5}.
\end{eqnarray}
The second step follows from the fact that the topology of $M^4$ is independent of the position on $I$, and so the integral of the characteristic class $Tr(R^2)$ is equal on both sides.  Thus the difference between the number of instantons on the two ends of the world is equal to the number of M5-branes in the bulk, as expected in the theory with Fabinger-Ho$\check{\textup{r}}$ava boundary conditions.

If we want to flip the chirality of the Weyl spinor on one of the sides, the correct boundary condition on that side is not Eq.~(\ref{const}) but rather its mirror image \cite{DFM}
\begin{equation}
G_4=-\frac{2Tr \tilde{F}^2+Tr R^2}{16\pi} \label{const2}
\end{equation}
where we have used the fact that parity is only an invariance of the 11$d$ SUGRA action if the sign of $G_4$ is flipped at the same time.  $\tilde{F}$ is the curvature of the $E_8$ gauge bundle after the action of this parity symmetry.  

If we wish to obtain the correct anomaly cancellation condition, we need to couple the opposite chirality fermions to the $E_8$ bundle with a parity flip in the bundle, which may be equivalent to a parity-flipped coupling to the original bundle.  This places a constraint on any proposed origin of the fermions, although perhaps the coupling of the opposite chirality spinor to the opposite parity bundle is not a very strong restriction.  Once (\ref{const2}) is imposed the known instanton relation for the Ho$\check{\textup{r}}$ava-Witten boundary conditions is automatic
\begin{eqnarray}
I_1+I_2&=&\frac{1}{16\pi^2}(\int_{(M^4,0)}Tr(F^2)+  \int_{(M^4,1)}Tr(\tilde{F}^2))\nonumber\\
&=&\frac{1}{2\pi}(\int_{(M^4,0)}G_4-\int_{(M^4,1)}G_4)-\int_{M^4}\frac{Tr(R^2)}{16\pi^2}\nonumber\\
&=&\frac{1}{2\pi}\int_{M^4\times I} dG_4-\int_{M^4}\frac{p_1(M^4)}{2}.
\end{eqnarray}
This identifies the sum of the instanton numbers as the usual factor that depends on the topology of $M^4$ minus the number of the instantons which have been pulled into the bulk, where they are M5-branes.

\section{\twoa\ Solitons from $LE_8$ and \twob\ from $LLE_8$} \label{twosec}
\subsection{Loopgroup Bundles from Dimensional Reduction}
The loopgroup of the group $G$, named $\Omega G$, is the group of closed loops in $G$ or equivalently the group of maps from the circle to $G$.  In this subsection we will review the fact \cite{Horava,slow} that configurations of $G$ bundles over $M\times S^1$ specify $\Omega G$ bundles over $M$, and so one may dimensionally reduce a $G$ gauge theory on a circle by replacing $G$ with its loopgroup, $\Omega G$.  The converse is not true, there are $\Omega G$ bundles over $M$ that cannot be lifted to $G$ bundles over $M\times S^1$.  In the current context we will see that such \twoa\ configurations which have no M-theory lift suffer from the Freed-Witten anomaly, as has been described for example in \cite{Manjarin,slow}.

We will be interested in a slightly more general case in which the original space-time is not necessarily a product $M\times S^1$ geometrically or topologically, but rather a circle bundle $Y$ over $M$.  In this case the loop group bundle must contain the data describing the connection of this circle bundle.  This is done by centrally extending the loopgroup by a circle.  Topologically this just means that our loopgroup $LG$ is the product of the free loopgroup described above with a circle
\begin{equation}
LG=\Omega G\times S^1.
\end{equation}
Intuitively this central-extension circle is just the original circle fibered over $M$, in fact the connection of this circle bundle is the same before and after the reduction.  We will later generalize this construction to allow the circle to be nontrivially fibered over the free loopgroup, and we will argue that in \twoa\ the Chern class of this nontrivial bundle is the Romans mass $G_0$.

To summarize, for every $G$ bundle $E$ over a circle bundle $Y$ over $M$ we want to use all of the data of both bundles to create an $LG$ bundle $F$ over $M$, where $LG$ is the trivially centrally-extended free loopgroup of $G$
\begin{equation}
\left\{\begin{array}{ccc}  G   &     \longrightarrow & E    \\
               &        & \dwn \\
           S^1 &     \longrightarrow& Y    \\
               &        & \dwn \\
               &        & M        \end{array}\right\}
~~~~~ \longrightarrow ~~~
\left\{\begin{array}{ccc}LG   & \!\! \longrightarrow & \!\! F    \\   
               &        & \!\! \dwn \\
               &        & \!\! M       \end{array}\right\}. \label{dimred}
\end{equation}
Both bundles may be constructed by trivializing over patches and then determining transition functions.  In the first case there are two types of transition functions, one for each of the bundles:
\begin{equation}
f:U\times S^1\longrightarrow G \hsp{.3} \textup{and} \hsp{.3} g:U\longrightarrow S^1
\end{equation}
where $U$ is the intersection of two local patches of $M$, where the transition functions are defined.  From these two functions we must make the transition functions of the second bundle
\begin{equation}
h:U\longrightarrow LG=\Omega G\times S^1.
\end{equation}
The transition function $h$ is just the direct sum of two functions, one from $U$ to $\Omega G$ and another from $U$ to $S^1$.  These may be defined from the pair $(f,g)$ as follows
\begin{equation}
h(x\in U)(\theta\in S^1)=(f(x,\theta),g(x))\in G\times S^1
\end{equation}
where we have used the fact that an element of the loopgroup $\Omega G$ is defined to be a function from $S^1$ to $G$ to find the image of a point $\theta\in S^1$.  Notice that the transition of the original circle bundle over $M$ in the unreduced picture is the same as the transition function of the central extension circle bundle in the reduced picture, in accordance with the intuition that the central extension circle is indeed the original circle that has been reduced away.  Any two distinct $G$ bundles over circles on the left side  of (\ref{dimred}) yield two distinct $LG$ bundles on the right side, and so distinct solitons in the $G$ gauge theory on a circle bundle over $M$ will yield distinct solitons on the dimensionally reduced $LG$ gauge theory on $M$.  

Loopgroups are in general infinite-dimensional.  But this does not imply that there will be an infinite number of massless gauge bosons.  On the contrary, the photons corresponding to nontrivial loops will acquire masses by the usual Kaluza-Klein mechanism because they correspond to derivatives with respect to the reduced circle.  Therefore the $LG$ gauge group will be Higgsed at least to $G$.  When the radius of the circle becomes infinite the Kaluza-Klein states become massless and here an infinite-dimensional gauge symmetry may be restored, reflecting the fact that the dimensional reduction is not valid at infinite radius.  Solitons that use the topology of the full unbroken $LG$ gauge group require that the full $LG$ symmetry is restored in their core, and thus the radius will need to be infinite in the core of such a soliton.  Indeed this is the case for the NS5-brane, the only soliton which will require the topology of the full $LE_8$.

\subsection{Solitons From Dimensional Reduction}
As has been reviewed in Subsec.~\ref{hom}, to get the soliton spectrum of a $G$ gauge theory one needs to know the homotopy groups of $G$ or equivalently the characteristic classes.  To get the solitons of the dimensionally reduced theory we therefore need the homotopy groups of the loopgroup.  These are
\begin{equation}
\pi_1(LG)=\pi_2(G)\oplus\pi_1(G)\oplus\Z\sp
\pi_{k\geq 2}(LG)=\pi_{k+1}(G)\oplus\pi_k(G). \label{hgs}
\end{equation}
This means that if the $\pi_k(G)=\Z$ then we will find a $\Z$ factor in $\pi_k(LG)$ and also $\pi_{k-1}(LG)$.  However $\pi_1(LG)$ has an extra $\Z$ coming from the circle by which the loopgroup is centrally extended.  Intuitively this is just the original circle on which we have dimensionally reduced, and so it must be kept to assure for example that the Kaluza-Klein monopoles of the dimensionally reduced circle may be found in the dimensionally reduced description.  

Again we may formally consider this theory to be some kind of dimensional reduction of the total space of the $G$ or $LG$ bundle, in which case one may formally define the size of the minimal cycles in $G$ and $LG$ which are the images of these homotopy groups.  If the volume of the minimal $k$-cycle in $G$ which is the image of $\pi_k(G)$ is equal to $V$, and the radius of the circle if $R$, then we define the ``volumes'' of the $1$, $k-1$ and $k$ cycles in $LG$ to be $R$, $V/R$ and $V$ respectively.

We recall that if $\pi_k(G)=\Z$ then there is a characteristic class $G_{k+1}$ in the $G$ gauge theory which describes this part of the topology of the $G$ bundle.  After dimensionally reducing this nontrivial homotopy class becomes $\pi_k(LG)$ and $\pi_{k-1}(LG)$ which yield characteristic classes $G_{k+1}$ and $G_k$ in the dimensionally reduced theory.  Of course this is what we expect from dimensional reduction, the original $G_{k+1}$ is a $k+1$-form and so each term may have a component along the circle direction or it may not, and in these two cases it reduces to a $k$-form or a $k+1$-form respectively.  In addition the extra factor in $\pi_1(LG)$ from the central extension leads to a 2-form characteristic class $G_2$, which is just the field strength for the Kaluza-Klein $U(1)$ gauge field.  Thus the homotopy groups of $LG$ are exactly what they need to be to reproduce the known field content of the KK-reduced theory.

One may reproduce the same results in a third way, by considering the effect of dimensional reduction upon the solitons of the $G$ gauge theory.  We have seen that in the unreduced theory $\pi_k(G)=\Z$ leads to magnetic defects of codimension $k+2$ and electrically charged objects of dimension $k$.  In the reduced theory the corresponding homotopy groups are $\pi_k(LG)=\pi_{k-1}(LG)=\pi_1(LG)=\Z$.  These lead to magnetic objects of codimensions $k+2$, $k+1$ and $3$ and also electric objects of dimensions $k$, $k-1$ and $1$.  As can easily be seen by comparison with the KK-reduction of the fields sourced by these solitons, the first two magnetic objects in the reduced gauge theory lift to the unreduced magnetic object unwrapped about the circle or wrapped respectively.  The third magnetic object in the reduced theory is the KK-monopole for the circle bundle.  Similarly of the three electric solitons in the reduced theory one comes from the unwrapped unreduced electric soliton, one from the wrapped unreduced electric soliton and the last is a Kaluza-Klein momentum mode for the compactified circle.  

In general the values of the tensions of these objects depend on the details of the gauge theory.  However the relations betweens a soliton and its dimensional reductions are straightforward.  The unwrapped reductions have the original tension, the wrapped reductions have a tension which is higher by the radius $R$ of the compactified circle, and the KK momentum modes are particles of mass $1/R$.  

In the case of M-theory there is only one length scale, which we use to define the ``size'' of the minimal 3-cycle in the $E_8$ fibers.  This scale is the inverse of the membrane tension, and so in M-theory the electric solitons have a tension equal to inverse volume of the cycle with which they are made.  The magnetic solitons, M5-branes, have a tension equal to the size of the minimal cycle, with a factor of $l_p^{-9}$ to fix the dimensions.  Thus the product of the electric and magnetic tensions is the constant $l_p^{-9}$.  Every time that we reduce on a circle of radius $R$, the resulting electric and magnetic pairs consist of one of the original solitons wrapped on the circle and its dual not wrapped.  Thus the product of the masses increases by the radius of the circle.  The original minimal $k$-cycle becomes a $k$-cycle and a $k-1$ cycle whose volume is $1/R$ of the volume of the $k$-cycle.  This ensures that in dimensional reductions of M-theory on torii the electric solitons each have a tension equal to the inverse of the size of the minimal cycle with which the soliton is made.  Similarly the magnetic cycles have a tension equal to the size of the minimal cycle multiplied by a factor of $l_p^{-9} V$, where $V$ is the volume of the torus.  These relations will allow us to easily relate soliton tensions with the sizes of the cycles with which they are formed.

\subsection{The Solitons of \twoa}

Now we will use the above technology to reproduce the field content and soliton spectrum of type \twoa\ supergravity.  It will come as no surprise that the correct spectrum is found, after all, so far the bundles appear to be little more than bookkeeping devices for the supergravity field content.  The less trivial result, T-duality, will appear in the next subsection.

We recall that M-theory solitons are classified by $E_8$ bundles over an 11-dimensional space-time which have a characteristic class $G_4$ describing transition functions inhabiting $\pi_3(E_8)=\Z$.  The volume of a minimal $SU(2)\subset E_8$ is $l_p^3$ and correspondingly the electric M2-brane has tension $l_p^{-3}$ while the magnetic M5 has a tension of $l_p^{-6}$.  

If we consider an 11-dimensional space-time which is the product of a 10-manifold $M^{10}$ and the M-theory circle $S^1_M$ of radius $R$ then we may use the data of any $E_8$ bundle over the 11-manifold to construct a $LE_8$ bundle over $M^{10}$.  Using the theorem (\ref{hgs}) we may easily calculate the low-dimensional homotopy groups of $LE_8$ to be
\begin{equation}
\pi_3(LE_8)=\pi_2(LE_8)=\pi_1(LE_8)=\Z.
\end{equation}
Using the above definition of the volume of the original $E_8$ and the recursive definitions for the volume of the loopgroup fibers, we find that the volumes of these minimal cycles are $l_p^3$, $\alpha\p=l_p^3/R$ and $R$ respectively.  The corresponding characteristic classes are the type \twoa\ field strengths $G_4$, $H$ and $G_2$.  These are the quantized, closed field strengths constructed by taking the exterior derivatives of the corresponding connections $G_p=dC_{p-1}$ on each patch.  The gauge invariant field strengths $dC_{p-1}+H\wedge C_{p-3}$ are unsuitable as they are in general neither quantized nor closed.  The magnetic solitons are the D4, NS5 and D6-branes while the respective electric solitons are the D2, F-string and D0-brane.  The D8-brane is absent, as predicted in \cite{Hull} these will appear once we compactify on (twisted) three-torii and will correspond to domain walls between different central extensions of $LE_8$.  

The tensions of the electric solitons are the inverses of the sizes of the cycle with which they are charged.  In the case of the D0-brane the tension $1/R$ is the expected tension of a KK momentum mode with respect to the compactified circle.  Similarly the tensions of the F-string and D2-brane may be reexpressed in the familiar forms $1/\alpha\p$ and $1/({\alpha\p}^{3/2}g_s)=l_p^{-3}$.
\begin{table}
\begin{center}
\begin{tabular}{c|c|c|c|c|c|c}
Homotopy&Volume&Char.&Electric&Elec. Sol.&Magnetic&Mag.Sol.\\
Class&of Cycle&Class&Soliton&Tension&Soliton&Tension\\
\hline
$\pi_3(LE_8)=\Z$&$l_p^3$&$G_4$&D2-brane&$1/l_p^3=1/({\alpha\p}^{3/2}g_s)$&D4&$1/({\alpha\p}^{5/2}g_s)$\\$\pi_2(LE_8)=\Z$&$\alpha\p$&$H$&F-String&$1/\alpha\p=R/l_p^3$&NS5&$1/({\alpha\p}^{3}g_s^2)$\\
$\pi_1(LE_8)=\Z$&$R$&$G_2$&D0-brane&$1/R=1/({\alpha\p}^{1/2}g_s)$&D6&$1/({\alpha\p}^{7/2}g_s)$\\
\end{tabular}
\end{center}
\caption{The soliton spectrum of \twoa\ SUGRA from $LE_8$ bundles}
\end{table}

As in the case of M-theory, the electric monopoles do not need to be included by hand, but may be constructed topologically from the construction of dielectric magnetic monopoles of higher dimensions wrapped about trivial cycles.  In several cases this construction is just the dimensional reduction of the above construction of M2-branes from dielectric M5-branes.  This exercise demonstrates again that the topology of the bundles seems to automatically account for the Chern-Simons terms which make such dielectric charges possible.  More generally the existence of this bundle may well be equivalent to the vanishing of the Freed-Witten anomaly \cite{Sethi,Manjarin,slow}.

The presence of an $LE_8$ bundle does not imply that there is an unbroken $LE_8$ gauge symmetry, but as mentioned above the nonzero-mode photons correspond to states in M-theory which are not constant with respect to the M-theory circle and so become massive by the Kaluza-Klein mechanism, leaving at most $E_8$ unHiggsed.  However the full generator of $\pi_2(LE_8)$ is used in the transition function on a sphere linking the NS5-brane, and so the corresponding 2-sphere of gauge symmetries is apparently unbroken in an NS5-brane core.  This is consistent with the fact that $R$ is infinite in the core of an NS5-brane, and so in the 10$d$ perspective the full infinite-dimensional gauge symmetry is restored.

\subsection{M-Theory on the Two-Torus}

This procedure may be iterated to compactify M-theory on a 2-torus $S^1_M\times S^1_{\twoa}$, where the radii of the two circles are $R_M$ and $R_{\twoa}$ respectively. $E_8$ solitons on $M^9\times S^1_M\times S^1_{\twoa}$ may be mapped to $LE_8$ solitons on $M^9\times S^1_{\twoa}$ which finally yield $LLE_8$ solitons on $M^9$.  $LLE_8$ is the group of maps from $S^1_{\twoa}$ to maps from $S^1_M$ to $E_8$ or equivalently $LLE_8$ is the group of maps from $S^1_M\times S^1_{\twoa}$ to $E_8$.  The weak homotopy type of $LLE_8$ may be found by applying the theorem (\ref{hgs}) to the group $G=LE_8$.  The low dimensional homotopy groups are
\begin{equation}
\pi_3(LLE_8)=\Z\sp
\pi_2(LLE_8)=\Z^2\sp
\pi_1(LLE_8)=\Z^3
\end{equation}
where $\pi_3$ is generated by the $E_8\subset LLE_8$.  The two generators of $\pi_2(LLE_8)$ are the images of $\pi_2(LE_8)$ under the two embeddings of $LE_8\subset LLE_8$ given by $LE_8=\{ S^1_M\longrightarrow E_8\}$ and $LE_8=\{ S^1_{\twoa}\longrightarrow E_8\}$.  $\pi_1(LLE_8)$ is generated by the three circles $S^1_M$, $S^1_{\twoa}$ and $S^1_{E_8}$, where $S^1_{E_8}$ is constructed from maps from the two-torus to an $SU(2)$ in $E_8$ or equivalently from $S^1_{\twoa}$ to $LE_8$.  

We note in passing that the irrelevance of the higher homotopy groups of $E_8$ to the considerations of this paper implies that the based part of $LE_8$ approximates $K(\Z,2)$.  $K(\Z,2)$ is defined to be the manifold such that $\pi_2(K(\Z,2))=\Z$ while all other homotopy groups vanish.  It is the classifying space for complex line bundles, and thus maps from $S^1_{\twoa}$ to $LE_8$ approximately yield line bundles over $S^1_{\twoa}$.  Using this perspective the minimal line bundles, which are parametrized by $S^1_{E_8}$, are those which have flat connections.  Thus we may interpret $S^1_{E_8}$ as the space of flat connections on a complex line bundle over $S^1_{\twoa}$.  Although we will not make extensive use of this perspective in the sequel, we note that this interpretation of $S^1_{E_8}$ as the Fourier-Mukai transform of $S^1_{\twoa}$ agrees with the description of T-duality in, for example, \cite{KentaroIndexTDuality}.

The $S^1_{E_8}$ bundle over $M^9$ is topologically characterized by a 2-form characteristic class, which is just the curvature $F$ of the circle bundle.  This two-form characteristic class arises from the dimensional reduction of the three-form characteristic class $H$ which describes the fibration of the nontrivial $S^2$ in $LE_8$ over $M^9\times S^1_{\twoa}$ in the 10-dimensional description.  Therefore $F$ may be found by integrating $H$ over $S^1_{\twoa}$
\begin{equation}
F=\int_{S^1_{\twoa}}H. \label{kurv}
\end{equation}

We may calculate the radius of $S^1_{E_8}=\{S^1_{\twoa}\longrightarrow LE_8\}$, which we recall is the inverse mass of its electric solitons, as we have calculated the volume of the 2-cycle in the previous subsection.  The radius $R_{E_8}$ is by definition just the volume $\alpha\p$ of the two-cycle in $LE_8$ divided by the radius $R_{\twoa}$ of $S^1_{\twoa}$.  Equivalently it is the volume $l_p^3$ of the minimal three-cycle in $E_8$ divided by the volume of the 2-torus $R_MR_{\twoa}$
\begin{equation}
R_{E_8}=\frac{\alpha\p}{R_{\twoa}}=\frac{l_p^3}{R_MR_{\twoa}}. \label{rad}
\end{equation}
These two equivalent expressions for $R_{E_8}$ correspond to the two descriptions of its electric soliton as an F-string wrapped on $S^1_{\twoa}$ and as an M2-brane wrapped on $S^1_M\times S^1_{\twoa}$. 

In type \twoa\ compactified on the circle $S^1_{\twoa} $ we consider $S^1_{E_8}$ to be an internal direction.  Although conservatively we may think of the $E_8$ bundle as little more than a bookkeeping device, we cannot deny that the Kaluza-Klein modes (electric solitons) of $S^1_{E_8}$ are very real.  As described above, they are winding modes of fundamental strings about $S^1_{\twoa}$ with a mass of $R_{\twoa}/\alpha\p$.  There is of course a full infinite tower of Kaluza-Klein modes, but for $R_{\twoa}>R_{E_8}$ the contributions from higher terms in the tower are suppressed by powers of the ratio of the radii and so this dimensional reduction can be sensible.  However when $R_{E_8}>R_{\twoa}$ the higher modes have divergent contributions to observables and the dimensional reduction is apparently ill-defined.  By Eq.~(\ref{rad}) this occurs precisely when $R_{\twoa}<\sqrt{\alpha\p}$, which is when the type \twoa\ description breaks down in the $\alpha\p$ expansion of perturbative string theory.  

Our proposed prescription for making sense of this theory is, as one would do in field theory, to consider $S^1_{\twoa}$ to be the internal direction and $S^1_{E_8}$ to be the ``physical'' direction when the latter is larger.

\noindent
\textbf{{\textit{Proposal:} \twob\ string theory configurations on an $S^1_{\twob}$ bundle over $M^9$ are classified by $LLE_8$ bundles over $M^9$, where $S^1_{\twob}$ is the $E_8$ circle $S^1_{E_8}\subset LLE_8$ described above.  These $LLE_8$ bundles may lift to $E_8$ bundles over $S^1_{M}\times S^1_{\twoa}$ bundles over $M^9$.  We define this $S^1_M\times S^1_{\twoa}$ to be the F-theory torus.}}

Intuitively from the 9-dimensional $LLE_8$ bundle we choose to dimensionally ``unreduce'' the larger of the circles $S^1_{\twoa}$ and $S^1_{E_8}$.  We propose that the unreduction of $S^1_{E_8}$ yields \twob\ string theory compactified on $S^1_{E_8}$, which we will henceforth often refer to as $S^1_{\twob}$.  We may then define the F-theory torus to be the $S^1_M\times S^1_{\twoa}$ fiber above the \twob\ 10-manifold, which is a $S^1_{\twob}$ bundle over $M^9$.  We note that the choice of F-theory torus is not uniquely determined by a \twob\ configuration, but one must also choose which circle in the \twob\ configuration is $S^1_{\twob}$.  We will see an example of this ambiguity later when we compactify M-theory on a twisted 3-torus, which will yield two circles in \twob.  In this case one of the choices will yield a T-duality to Romans \twoa\ and so there will be no 11d lift, however even in this case the F-Theory torus may still be constructed from $S^1_{\twoa}$ and the central extension of $LE_8$.

As a simple check of our proposal the soliton spectrum is that of \twoa\ on a circle and so automatically agrees with that of \twob\ on a circle.  In addition we have already seen that KK modes with respect to the \twob\ circle are F-string winding modes with respect to the \twoa\ circle.  The radii of the two circles (\ref{rad}) agrees with the usual relation for T-dual circles, and less trivially we have seen (\ref{kurv}) that the curvature of the $S^1_{\twob}$ bundle in the space-time of \twob\ is equal to the integral of the $H$ flux in type \twoa\ integrated over $S^1_{\twoa}$.  In Ref.~\cite{BEM} it has been argued that this is indeed the correct formula for the topology change when one T-dualizes a circle that supports $H$-flux.  The cases in which $S^1_{\twoa}$ are trivially fibered have been understood since \cite{AABL}, however with a little more work we may trace the curvature $F_A$ of the $S^1_{\twoa}$ bundle to the dimensional reduction of $H_B$ in \twob\ and thus find
\begin{equation}
F_A=\int_{S^1_{\twob}}H_B \label{kurvb}
\end{equation}
in agreement with the general case in Ref.~\cite{BEM}.

\section{Examples} \label{exsec}

To further test the identification of the circle $S^1_{E_8}\subset LLE_8$ of a \twoa\ compactification with the circle $S^1_{\twob}$ of the T-dual \twob\ compactification we will consider several examples.  In this section we will consider two \twoa\ compactifications on $M^9\times S^1_{\twoa}$, which correspond to $LLE_8$ bundles over $M^9$.  We will see that the $S^1_{E_8}$ subbundle of this $LLE_8$ is the expected \twob\ space-time.  In the second example we will see that if we associate $S^1_{\twoa}\times S^1_M$ with the F-theory torus we further obtain the correct monodromy about the D7-brane. 

\subsection{NS5s to KK-Monopoles}

An NS5-brane is characterized by the fact that a 3-sphere that links it is the base of a nontrivial $LE_8$ bundle, where the transition function on the 2-sphere equator is the generator of $\pi_2(LE_8)=\Z$.  Of course an NS5-brane can be linked by any number of different 3-manifolds, each of which comes with a nontrivial $LE_8$ bundle.  In general this $LE_8$ bundle can be found by constructing a projection map from the three-manifold to the sphere which preserves the linking number of the NS5-brane and pulling back the nontrivial bundle on the 3-sphere by this map\footnote{This is because the bundle is characterized by the integral of $H$ over the 3-manifold, and due to Gauss' Law and the fact that the linking number with the NS5-brane is equal to one this can be taken to be the generator of the integral third cohomology of the 3-manifold.  This generator is the image of the generator of the third cohomology of $S^3$ if the projection is degree one.}.

\begin{figure}[ht]
  \centering \includegraphics[width=5in]{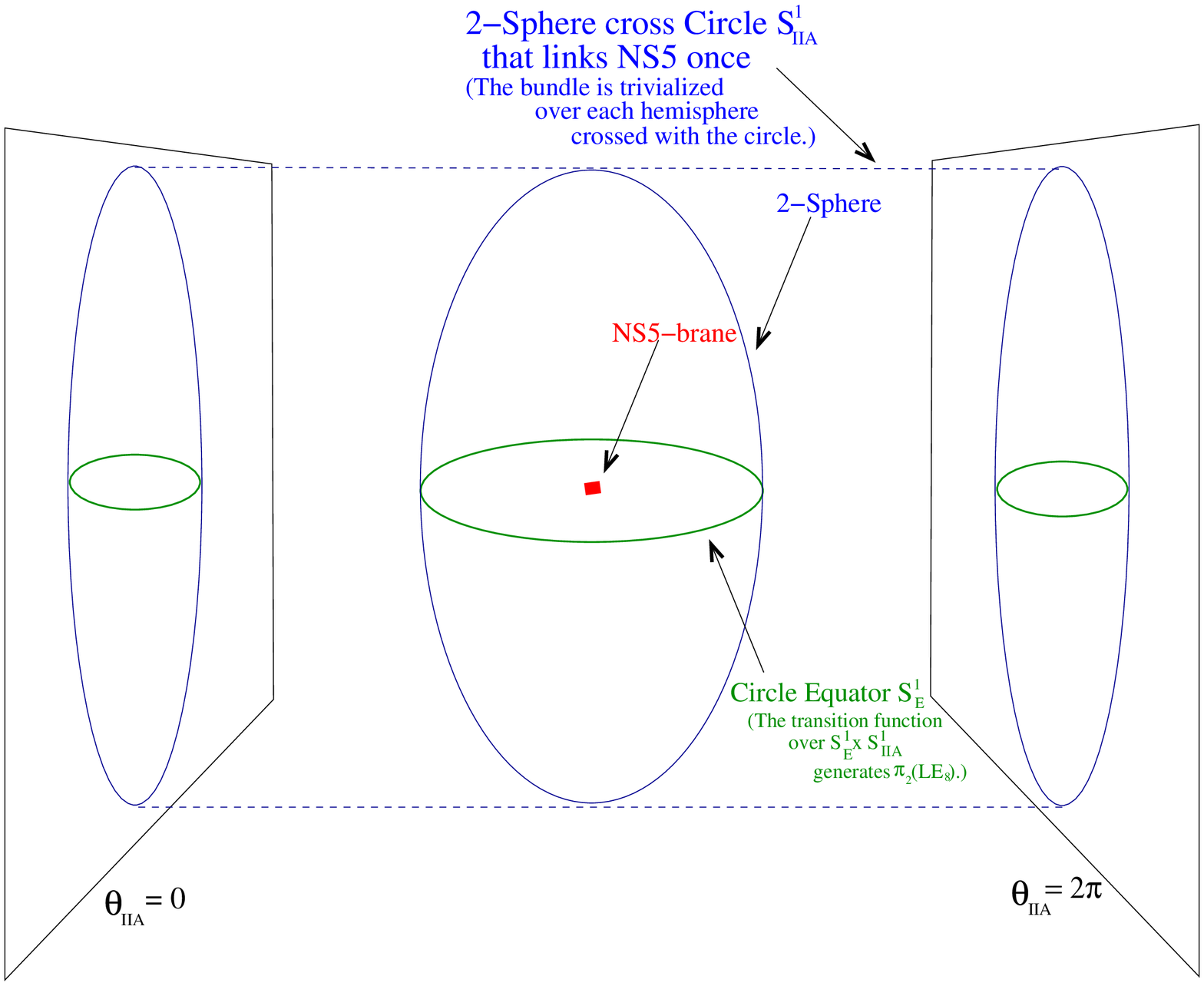}
\caption{\twoa\ string theory is compactified on $M^9\times S^1_{\twoa}$.  An NS5-brane is located at a fixed point $\theta_{\twoa}$ in the $S^1_{\twoa}$ direction and is linked by $S^2\times S^1_{\twoa}$, which is the base of a nontrivial $LE_8$ bundle. The transition function $S^1_E\times S^1_{\twoa}\longrightarrow LE_8$ is nontrivial and defines the transition function $S^1_E\longrightarrow S^1_{\twob}\subset LLE_8=\{S^1_{\twoa}\longrightarrow LE_8\}$ of a fibration of the \twob\ circle.  The nontrivial $S^1_{\twob}$ bundle is a KK-monopole configuration in the \twob\ description.} \label{ns5kk}
\end{figure}

For example if \twoa\ is compactified on $M^9\times S^1_{\twoa}$ one may consider an $S^2\times S^1_{\twoa}$ that links the NS5, as drawn in Fig.~\ref{ns5kk}.  A choice of projection map is the one used in the construction of the 3-sphere as the smash product of a one and two-sphere.  In accordance with Gauss' law, pulling back $H$ or equivalently pulling back the $LE_8$ bundle by this projection map we find that $H$ is the generator of the third cohomology of $S^2\times S^1_{\twoa}$ and so this is the unit nontrivial $LE_8$ bundle.  To construct this bundle we may cut $S^2\times S^1_{\twoa}$ into two patches $S^2_N\times S^1_{\twoa}$ and $S^2_S\times S^1_{\twoa}$.  The transition function is a map $f:S^1_E\times S^1_{\twoa}\longrightarrow LE_8$, where $S^1_E$ is the equator of the $S^2$.  Using the above smash product projection, we see that the transition function $f$ can be constructed by composing the smash product map $S^1_E\times S^1_{\twoa}\longrightarrow S^2$ with the generator of $\pi_2(LE_8)=\Z$ which is a map $S^2\longrightarrow LE_8$.

In the 9-dimensional perspective this configuration is described by an $LLE_8=\{S^1_{\twoa}\longrightarrow LE_8\}$ bundle over $M^9$.  The $S^2\times S^1_{\twoa}$ is dimensionally reduced to $S^2$, and so the nontrivial bundle resides over this $S^2$.  The $S^2$ may again be trivialized on each hemisphere, and the transition function is $g:S^1_E\longrightarrow \{S^1_{\twoa}\longrightarrow LE_8\}$. $g(\theta_E\in S^1_E)$ is therefore defined by the image of each point $\theta_{\twoa}\in S^1_{\twoa}$
\begin{equation}
g(\theta_E\in S^1_E)(\theta_{\twoa}\in S^1_{\twoa})=f(\theta_E,\theta_{\twoa}). \label{iso}
\end{equation}
The non-triviality of $f$ leads to the non-triviality of $g$, as Eq.~(\ref{iso}) is the isomorphism of maps
\begin{equation}
\{S^1_E\longrightarrow LLE_8\}=\{S^1_E\longrightarrow \{S^1_{\twoa}\longrightarrow LE_8\}\}\cong \{S^1_E\times S^1_{\twoa}\longrightarrow LE_8\}.
\end{equation}

We may now ``unreduce'' $S^1_{\twob}$.  This means that $S^1_{\twob}$ is now considered to be a space-time direction.  It is nontrivially fibered over 2-spheres that link the submanifold of $M^9$ formerly occupied by the NS5-brane, and so the space-time circle $S^1_{\twob}$ is nontrivially fibered over such 2-spheres.  In fact the bundle over each 2-sphere is the just Hopf bundle, as we could have deduced immediately from Eq.~(\ref{kurv}) and Gauss' law
\begin{equation}
\int_{S^2} F=\int_{S^2\times S^1_{\twoa}}H=1.
\end{equation}  

The fact that the circle bundle is nontrivial over a contractible 2-sphere in $M^9$ means that it degenerates somewhere inside every 2-sphere, and so on the submanifold where the NS5-brane was in \twoa.  The degeneration of the circle bundle just described is a Kaluza-Klein monopole with respect to the circle $S^1_{\twob}$, which is the expected T-dual of an NS5-brane.

\subsection{D6s to D7s and F-Theory}

\begin{figure}[ht]
  \centering \includegraphics[width=4.5in]{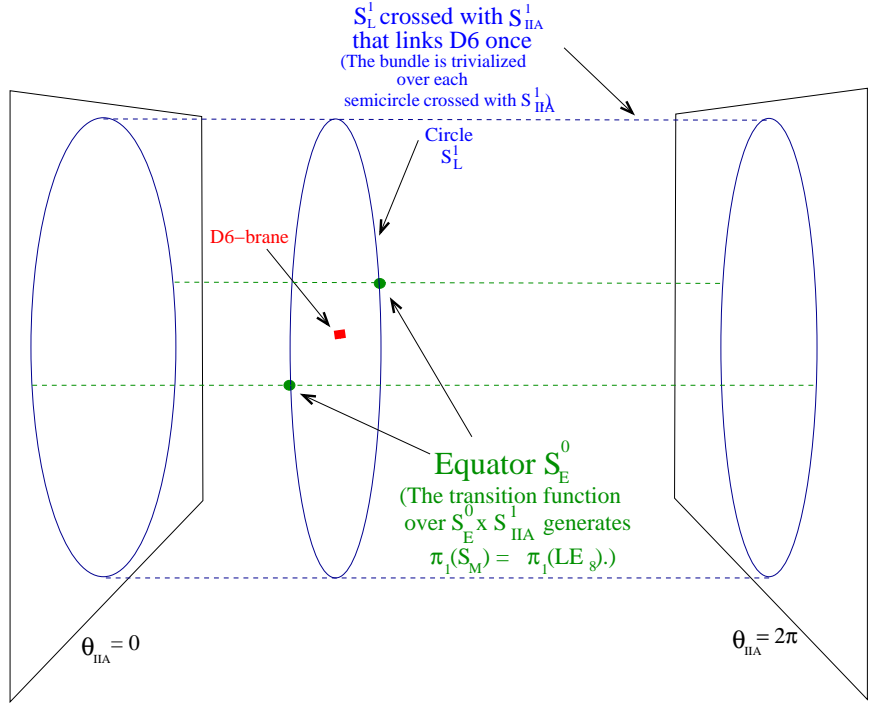}
\caption{\twoa\ string theory is compactified on $M^9\times S^1_{\twoa}$.  A D6-brane is located at a fixed point $\theta_{\twoa}$ in the $S^1_{\twoa}$ direction and is linked by $S^1_L\times S^1_{\twoa}$, which is the base of a nontrivial $S^1_M\subset LE_8$ bundle. At one of the two points of the equator $S^0_E$ the transition function $S^1_{\twoa}\longrightarrow S^1_M$ is $\theta_M\mapsto\theta_M+\theta_{\twoa}$.  This is a $T$-type transformation in the $SL(2,\Z)$ action on the 2-torus $S^1_M\times S^1_{\twoa}$.  In the \twob\ description $S^1_L$ is a loop about a D7-brane and the transition function is the monodromy about the loop, which is the expected $SL(2,\Z)$ action on the F-theory torus $S^1_M\times S^1_{\twoa}$.} \label{d6d7}
\end{figure}

We have claimed not only that unreducing the $E_8$ circle yields \twob, but further that the torus $S^1_M\times S^1_{\twoa}$ bundle over the \twob\ configuration provides the F-theory torus fibration.  As a test we will check that this torus has the expected monodromy as one encircles a D7-brane.  The D7-brane, in turn, will be constructed by T-dualizing a D6-brane with respect to a transverse circle $S^1_{\twoa}$, as in Figure~\ref{d6d7}.

We begin again with type \twoa\ on $M^9\times S^1_{\twoa}$, this time with a D6-brane that is located at a point $\theta_{\twoa}\in S^1_{\twoa}$.  For simplicity we do not include NSNS flux.  The D6-brane is defined by the fact that any 2-sphere linking the D6-brane is the base of a nontrivial $S^1_M\subset LE_8$ bundle, the Hopf bundle.  As above, we will be interested in not a 2-sphere linking it, but rather the 2-torus $S^1_L\times S^1_{\twoa}$.  This torus also supports a nontrivial $S^1_M$ bundle, the Poincare bundle.  

We may trivialize the Poincare bundle on $S^1_N\times S^1_{\twoa}$ and $S^1_S\times S^1_{\twoa}$, where $S^1_N$ and $S^1_S$ are the northern and southern semicircles of $S^1_L$.  If we define $S^0_E$ to be the equator of $S^1_L$, which consists of two points, then the transition function is a map $f:S^0_E\times S^1_{\twoa}\longrightarrow S^1_M$.  Only the monodromy about $S^1_L$ affects the topology of the bundle, and so without changing the bundle's topology we may set the transition function at one point in $S^0_E$ equal to the composition of the transition functions at both points in $S^0_E$, and then set the function at the other point to the identity.  Thus this bundle is determined by a single function, which being the generator of $\pi_1(S^1_M)$ must be
\begin{equation}
f:S^1_{\twoa}\longrightarrow S^1_M:\theta_{\twoa}\mapsto\theta_M.
\end{equation}
In other words each time one circumnavigates $S^1_L$ the transition function augments the M-theory position $\theta_M$ by the \twoa\ position $\theta_{\twoa}$.  This is just the $T$-transformation
\begin{equation}
T=\left(\begin{matrix} 1&1\\0&1\\
\end{matrix}\right)
\in SL(2,\Z) \label{t}\end{equation} 
in the $SL(2,\Z)$ action on the torus $S^1_M\times S^1_{\twoa}$.

We may now dimensionally-reduce away $S^1_{\twoa}$, and so this torus becomes a subset of the $LLE_8$ fibered over $M^9$.  The dimensionally reduced D6-brane is still 6+1-dimensional, and so it is now linked by a circle.  The torus subbundle over this circle is nontrivial, the monodromy is the $T$-transformation (\ref{t}).  Finally we unreduce the circle $S^1_{\twob}\subset LLE_8$.  Using Eq.~(\ref{kurvb}) we find that the \twob\ space-time manifold is $M^9\times S^1_{\twob}$.  The circle $S^1_{\twob}$ was not involved in the above construction and, as there is no NSNS-flux, nothing depends on the $S^1_{\twob}$ position.  In particular the D6-brane extends in this new direction and so is a 7-brane.  The 2-subtorus $S^1_M\times S^1_{\twoa}$ may be pulled back to this \twob\ space-time, where we find that it is still nontrivially fibered.  The monodromy about any lift of the loop $S^1_L$, which links the D7-brane, is the $T$-transformation on the F-theory torus.  This identifies our 7-brane as the expected D7-brane.

\section{Massive \twoa\ from the Central Extension} \label{romsec}

\subsection{Hull's Proposal for Massive \twoa}

We have now seen that the bosonic parts of configurations in M-theory are classified by $E_8$ bundles over 11-dimensions.  Using the Ho$\check{\textup{r}}$ava-Witten construction we have extended this to the $E_8\times E_8$ heterotic theory.  Dimensionally reducing on a circle we have found a similar description for \twoa\ configurations in terms of $LE_8$ bundles.  Further dimensionally reducing produces a new circle, and unreducing on that yields the T-dual \twob\ or equivalently F-theory configurations.  Presumably some such strategy also allows us to find the Heterotic $SO(32)$ or equivalently type I configurations which are T-dual to heterotic $E_8\times E_8$ configurations.

This leaves at least two types of configurations to be found.  First, there may be type \twob\ or heterotic $SO(32)$ configurations which have no free circle action with respect to which we can T-dualize.  In particular a decompactification of $S^1_{\twob}$ is dual to the limit in which the size $l_p$ of the $E_8$ fiber becomes infinite, a limit which is difficult to interpret if it exists at all.  Of course there is always a nonfree circle action, and in fact the \twob\ circle in the KK monopole solution which is T-dual to the NS5-brane configuration degenerates and our construction seems to have mysteriously continued to work, but in a more general setting such as mirror symmetry there may be extra light degrees of freedom in such cases that need to be considered.

The second obvious omission thus far is massive \twoa.  As noted above, our \twoa\ configurations never have D8-branes.  Furthermore in massive \twoa\ D0-branes and NS5-branes are confined by F-strings and D6-branes respectively, but we have found no evidence of this in the \twoa\ reduction above, reflecting the fact that our $LE_8$ bundle configurations are massless \twoa\ configurations.  In Ref.~\cite{slow} the authors show that, using the realization of massive \twoa\ that we will find in this section, the confinement of NS5-branes is automatic and the number of D6-branes ending on each NS5-brane (the tension of the confining string) is, as desired, equal to the Romans mass.  

To obtain massive \twoa\ in our setting we will use Hull's proposal, which states that massive \twoa\ string theory is related by two 2 T-dualities to M-theory on a 3-dimensional nilmanifold.  This is quite reasonable as a 3-dimensional nilmanifold may be realized as a $S^1_M$ bundle over the 2-torus $S^1_{\twoa}\times S^1_{\twob}$ with Chern class $k$.  Here we have abused our notation by referring to a space-time circle, which is not the $E_8$ circle, as $S^1_{\twob}$.  This is because it will become the $E_8$ circle in massive \twoa, $i.e.$ after two T-dualities.  Such a configuration, after dimensionally reducing on $S^1_M$, is \twoa\ compactified on $M^8\times S^1_{\twoa}\times S^1_{\twob}$ with
\begin{equation}
\int_{S^1_{\twoa}\times S^1_{\twob}}G_2=2\pi k.
\end{equation}
T-dualizing on $S^1_{\twoa}$ yields a \twob\ configuration with $\int_{S^1_{\twob}}G_1=2\pi k$ and further T-dualizing with respect to $S^1_{\twob}$ yields a massive \twoa\ configuration with $G_0=2\pi k$.  From now on we will drop the factor of $2\pi$ in $G_0$.  The goal of the present section is to see what this chain of dualities does to our $E_8$ bundle, which will teach us the $E_8$ bundle realization of massive \twoa.

In fact we will show something slightly more general.  We will consider a configuration with D7-branes in \twob\ and T-dualize it to create D8-branes in \twoa.  We will consider boundary conditions such that one side of each brane carries flux and the other does not.  To obtain the above T-duality between \twob\ with $G_1$ flux and massive \twoa\ with $G_0$ flux one may simply restrict attention to the side of the branes with the flux by moving the branes away to infinity on the fluxless side. 

\subsection{The Romans Mass is the Central Extension of $LE_8$}

\begin{figure}[ht]
  \centering \includegraphics[width=4.5in]{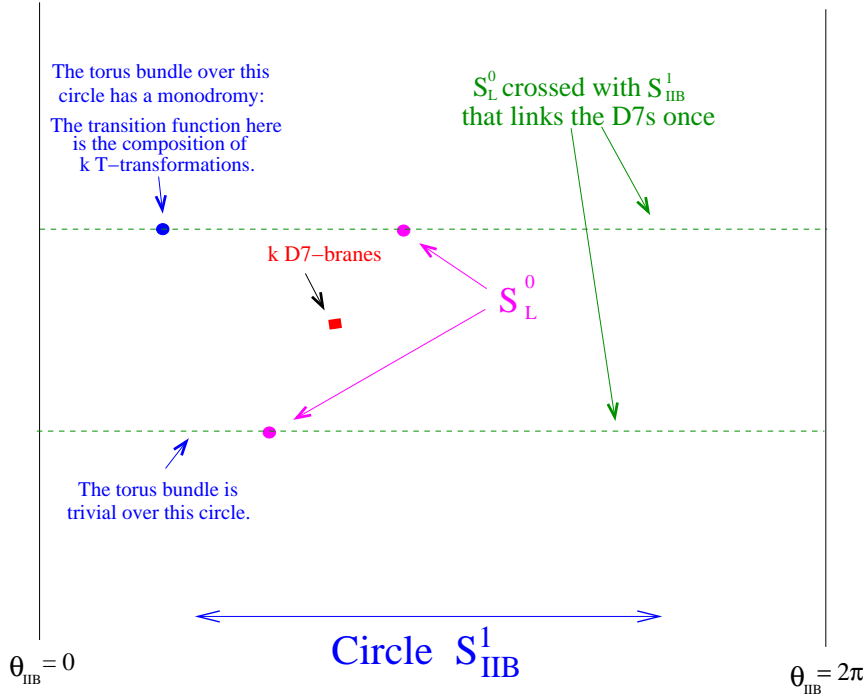}
\caption{\twob\ string theory is compactified on $M^9\times S^1_{\twob}$.  $k$ D7-branes are located at a fixed point $\theta_{\twob}$ in the $S^1_{\twob}$ direction and are linked by the two circles $S^0_L\times S^1_{\twob}$, which are the base of a nontrivial $S^1_M\times S^1_{\twoa}$ F-theory torus bundle.  If there is a trivial bundle over one of these circles then the other must support a nontrivial bundle, where the monodromy is the composition of $k$ $T$-transformations in $SL(2,\Z)$.  Reducing away the bundle's base $S^1_{\twob}$, the base is interpreted as the $E_8$ circle $S^1_{E_8}$ in the $LLE_8$ fiber over $M^9$.  This is the same bundle we find by dimensionally reducing an $LE_8^k$ bundle over $M^9\times S^1_{\twoa}$, where $k$ is the central extension of the loop group.  We expect the T-dual to a D7-brane to be a D8-brane, and so $LE_8^k$ bundle configurations appear to classify massive \twoa\ solitons with Romans mass $G_0=k$.} \label{d7d8}
\end{figure}

The T-duality between a D6-brane in \twoa\ and D7-brane in \twob\ has already been described in Sec.~\ref{exsec}, and so we may begin already in \twob.  We consider type \twob\ compactified on $M^9\times S^1_{\twob}$, with $k$ D7-branes localized at a point $\theta_{\twob}\in S^1_{\twob}$.  If we wish to return to a D6-brane configuration in \twoa\ we will need to T-dualize along a circle that is wrapped by $k=1$ D7-branes, which is necessarily a circle action on the $M^9$.  The dimensional reduction to the 9-dimensional space of orbits of this circle action is the base space of an $LLE_8$ bundle, and the orbits of the circle are the fibers of the $E_8$ circle in this $LLE_8$.  This $LLE_8$ can be lifted to an $E_8$ bundle over an 11-dimensional torus bundle over $M^9$.

Instead we will be interested in a T-duality with respect to $S^1_{\twob}$, whose dimensional reduction will give a \textit{different} $LLE_8$ bundle from the aforementioned dimensional reduction with respect to a circle action on $M^9$.  This may lead in particular to a distinct F-theory realization of this \twob\ configuration, despite the fact that the monodromies of the torus are determined by the $G_1$ fluxes and therefore must be the same in any F-theory lift of a \twob\ configuration.  In particular whether a D3-brane ``wraps'' the F-theory torus appears to depend on whether the circle-action that we T-dualize preserves the embedding of the D3-brane.

The 2-torus bundle over any loop $S^1$ is characterized by the monodromy about that loop.  In backgrounds like ours where the dilaton is single-valued the monodromy is always some composition of $T$-transformations determined by the integral of the RR field strength $G_1$ over the loop
\begin{equation}
v\mapsto T^kv\sp
v=\left(\begin{matrix}\theta_M\\\theta_{\twoa}\end{matrix}\right)\sp
T=\left(\begin{matrix} 1&1\\0&1\\
\end{matrix}\right)
\in SL(2,\Z) \sp 
\int_{S^1}G_1=2\pi k.
\end{equation}
In our case, sketched in Fig.~\ref{d7d8}, the $k$ D7-branes are linked by $S^0_L\times S^1_{\twob}$, the disjoint union of two copies of $S^1_{\twob}$.  The bundle over the cylinder orthogonal to the D7-branes is then characterized by the $G_1$ flux integrated over these two circles.  These fluxes must both integrate to integers (times $2\pi$) or else the partition function of a D(-1)-brane or D-instanton would be ill-defined, corresponding to the fact that $T^k$ is an element of $SL(2,\Z)$ only when $k$ is integral.  

We will consider the simplest case, in which the integral of $G_1$ over one circle is trivial, yielding the trivial bundle.  The difference between the two integrals, by Gauss' Law, is equal to the D7-brane charge linked.  Thus $G_1$ integrated over the second circle is equal to $2\pi k$, providing a monodromy of $T^k$ as the F-theory torus is pushed around this loop.

Dimensionally reducing this configuration on $S^1_{\twob}$ we might hope to find an $LLE_8$ bundle over $M^9$, but in fact we find something slightly different.  As expected our fiber has three circles, the \twob\ circle which we will call $S^1_{E_8}$ and also the two circles from the F-theory torus which we call $S^1_M$ and $S^1_{\twoa}$.  However in contrast with the usual topology of $LLE_8$, the torus is nontrivially fibered over $S^1_{E_8}$ producing a nilmanifold.  In the dimensional reduction procedure that we have defined lower-dimensional cycles do not affect higher-dimensional cycles in the reduced theory, and so the two and three-cycles of $LLE_8$ appear to be in tact.

One might already guess that a nontrivial bundle over $S^1_{E_8}$ in the 9-dimensional picture leads to a nontrivial bundle over the two-sphere $S^2\subset LE_8$ from which it arises in the 10-dimensional type \twoa\ picture, where we have unreduced with respect to $S^1_{\twoa}$.  After all if we take the $S^1_{\twoa}$ from the torus fiber and reinterpret it, using the smash product $T^2\longrightarrow S^2$, as part of the base then we find precisely this bundle over the $S^1_{\twoa}\wedge S^1_{E_8}=S^2\subset LE_8$.  While ``unreducing'' may not be well-defined, we will support this conjecture by starting with a centrally extended $LE_8$ bundle over 10-dimensions and reduce to obtain the variant of $LLE_8$ just described.

We start in \twoa\ with a stack of $k$ 8-branes.  On one side of the 8-branes is an $LE_8$ bundle, on the other side is an $LE_8^k$ bundle where $k$ is the central extension of the affine $E_8$.  Topologically the Lie group $LE_8^k$ is identical to the trivially centrally-extended $LE_8=LE_8^{k=0}$ except that the central extension $S^1_M$ is nontrivially fibered over the nontrivial 2-sphere $S^2\subset LE_8$ with Chern class $k$.  The fibration of these bundles over the base will not be important for the present discussion, we will see that $G_0$ is determined solely by the internal structure of the $LE_8^k$ fiber.  Of course this is in accord with the fact that $G_0$ is a space-time zero-form.

The $S^1_M$ bundle over $S^2$ is the $k$th tensor power of the Hopf fibration.  Rather than treat this bundle directly, we will find it to be slightly easier to introduce an auxiliary bundle over the two-torus $S^1_{E_8}\times S^1_A$, also with Chern class equal to $k$.  The smash product $s:T^2\longrightarrow S^2$ provides a projection map from the torus onto the sphere and the bundle over the torus is the pullback under $s$ of the bundle over the sphere.  This is, not accidentally, the same strategy we used with the space-time torus when we dimensionally reduced a configuration with a D6-brane.

The bundle over the torus is the $k$th tensor power of the Poincare bundle, and so the construction of the space of flat connections $S^1_{E_8}$ of the $S^1_M$ bundle over $S^1_A$ is just the restriction map.  That is, for each $\theta_{E_8}\in S^1_{E_8}$ the corresponding flat connection is the subbundle over $(\theta_{E_8},S^1_A)\subset S^1_{E_8}\times S^1_A$.  In particular the point $\theta_{E_8}\in S^1_{E_8}\subset LLE_8$ corresponds to loop $(\theta_{E_8},S^1_A)\subset S^1_{E_8}\times S^1_A$ pushed forward by the smash product $s$ into $LE_8$.  To save notation the other coordinates in the $LE_8$ group manifold have been omitted in the preceding interpretation of the points $\theta_{E_8}$ in the $E_8$ circle $S^1_{E_8}$.

In particular as one encircles $S^1_{E_8}$, the loops $(\theta_{E_8},S^1_A)$ sweep out all of $S^1_{E_8}\times S^1_A$.  If one considers a section of the $S^1_M$ bundle over $(\theta_{E_8},S^1_A)$ and pushes $\theta_{E_8}$ around $S^1_{E_8}$ once one necessarily finds that this section now winds $k$ more times around $S^1_M$.  Thus the monodromy about $S^1_{E_8}$ is the action $T^k$ on the torus $S^1_M\times S^1_A$.  $S^1_A$ parametrizes the circle in the definition of the loop group $LLE_8$, and so parametrizes $S^1_{\twoa}$.  Thus it is the torus $S^1_M\times S^1_{\twoa}$ that undergoes the monodromy of $T^k$ as one circumnavigates $S^1_{E_8}$ in this variant of $LLE_8$.  

This is precisely the relation that characterized the variant of $LLE_8$ found above by dimensionally reducing a circle in type \twob\ that supports $G_1$ flux.  This identifies our variant of $LLE_8$ as $L(LE_8^k)$, and leads us to conjecture that $L(LE_8^k)$ bundles over 9-dimensions characterize massive \twoa\ compactified on a circle\footnote{One may also use (\ref{hgs}) to verify that it has the expected homotopy groups.}.  Unreducing the circle $S^1_{\twoa}$ we appear to find that configurations of uncompactified massive \twoa\ are classified by $LE_8^k$ bundles over space-time.  Such bundles have no obvious 11-dimensional lift, rather the 11th dimension $S^1_M$ is hopelessly entangled with the $E_8$ coordinates, being fibered over the $S^2\subset LE_8$.  However if there are two free circle actions available then such a configuration is related by dimensional reductions and unreductions to an $E_8$ bundle compactified on a twisted 3-torus.

\section{Ramblings}
We have missed the most interesting cases, T-duality on circles which are allowed to degenerate.  We hope that any insight this may give into mirror symmetry may allow for real calculations, which after all should be the goal of this program.  In particular the generalization of Eq.~(\ref{kurv}) to the case in which the circle bundles may degenerate is unknown.  A more straightforward project would be to generalize this construction to the T-duality between the heterotic theories and to type I and I$\p$, as the generalization of Eq.~(\ref{kurv}) is unknown even there.

Using the above definition of F-theory the torus consists of two space-like circles and so there is no second time.  This is not problematic as one does not need to suggest that the low energy description is a 12-dimensional supergravity.  On the contrary, as one of the F-theory directions is T-dual to a space-time direction the approach of Ref.~\cite{RV} may imply that to obtain a classical description one must choose a gauge, which appears to be an 11-dimensional projection of this 12-dimensional space.  Thus while supersymmetry and even Lorentz invariance may not be manifest in the 12-dimensional description, it is enough that they appear in our gauge choices.

The notion that minimal cycles in the fiber have a ``size'' may be a formal bookkeeping device to keep track of the tensions of the electrically charged solitons.  But the picture in which solitons are somehow lifted into the total space of the bundle does seem to reproduce the dielectric effects caused by worldvolume couplings.  This leads one to wonder whether worldvolume actions may thus somehow be written as Jacobian determinants of such embeddings.  In particular the M2-brane action, at least in superspace, is perhaps the simplest place to start.  However if one wishes to take these fiber directions seriously, it may be essential that some mechanism, generalizing the gauge-fixing described in the last paragraph, allows one to eliminate the gravitational degrees of freedom that might otherwise deform them.

\bigskip\bigskip

\noindent 
{\bf Acknowledgements}

\noindent  
I am grateful to A. Adams, A. Adem, M. Douglas, J. J. Manjarin and S. Sethi for dispensing illumination, and in particular to P. Ho$\check{\textup{r}}$ava for his insight into the dimensional reduction program at an early stage.  I am yet more grateful to the INFN for keeping me alive while I wrote this up.

\noindent


\end{document}